\documentclass[prl, superscriptaddress, twocolumn, 10pt]{revtex4}  
\usepackage{amssymb}
\usepackage{amsmath}
\usepackage{bbm}
\usepackage{graphicx}
\usepackage{color}

\begin{document}

\author{Philipp Werner}
\affiliation{Department of Physics, University of Fribourg, 1700 Fribourg, Switzerland}
\author{Shintaro Hoshino}
\affiliation{Department of Physics, Saitama University, Saitama 338-8570, Japan}

\title{Nickelate superconductors -- Multiorbital nature and spin freezing}

\date{\today}

\hyphenation{}

\begin{abstract} 
Superconductivity with a remarkably high $T_c$ has recently been found in Sr-doped NdNiO$_2$ thin films. While this system bears strong similarities to the cuprates, some differences, such as a weaker antiferromagnetic exchange coupling and possible high-spin moments on the doped Ni sites have been pointed out. Here, we investigate the effect of Hund coupling and crystal field splitting in a simple model system and argue that a multiorbital description of nickelate superconductors is warranted, especially in the strongly hole-doped regime. We then look at this system from the viewpoint of the spin-freezing theory of unconventional superconductivity, which provides a unified understanding of unconventional superconductivity in a broad range of compounds. Sr$_{0.2}$Nd$_{0.8}$NiO$_2$ falls into a parameter 
regime influenced by two spin-freezing crossovers, one related to the emergent multi-orbital nature in the strongly doped regime and the other related to the single-band character and square lattice geometry in the weakly doped regime. 
\end{abstract}


\maketitle 

Nickelate analogs of the cuprates such as LaNiO$_2$ had been theoretically proposed more than 20 years ago \cite{Anisimov1999}, but only very recently has superconductivity been found in Sr-doped NdNiO$_2$ thin films \cite{Li2019}. This exciting discovery offers a new platform to study unconventional superconductivity and may provide new insights into the pairing mechanism in cuprate-like systems. Several theoretical investigations on the new compound have already been conducted \cite{Jiang2019,Sakakibara2019,Hepting2019,Wu2019,Nomura2019}. They essentially confirm the results of earlier bandstructure calculations \cite{Lee2004}, which suggest an intrinsic hole-doping of the Ni 3$d_{x^2-y^2}$ band by Nd 5$d$ pockets. The presence of the 5$d$ states at the Fermi surface led to speculation about an important role of the hybridization between the strongly correlated 3$d$ and more extended 5$d$ states, and possible analogies to heavy-fermion superconductivity \cite{Li2019,Hepting2019}. A detailed ab-initio study however suggests an almost perfect decoupling between the Ni 3$d_{x^2-y^2}$ states and those in the Nd layer \cite{Nomura2019}. The close analogy to the cuprates and the relatively high $T_c\sim 10$ K, which cannot be explained by a phonon-mediated pairing mechanism \cite{Nomura2019}, suggests unconventional superconductivity with most likely a $d$-wave order parameter \cite{Sakakibara2019,Wu2019}.

Two potentially relevant differences between the nickelate and cuprate superconductors have however been pointed out \cite{Lee2004,Jiang2019}. One is the substantially smaller antiferromagnetic exchange coupling resulting from the larger splitting between the Ni 3$d$ and O $2p$ bands in the nickelates. This appears to pose a problem if one tries to explain high-$T_c$ superconductivity as a pairing induced by antiferromagnetic spin fluctuations. The other difference is that the nickelate compound is a (doped) Mott insulator, where the doped holes end up on the Ni sites, whereas the cuprates are classified as charge transfer insulators \cite{Zaanen1985}, where the doped holes are on the O sites. In the nickelates, there is hence a possibility of high-spin ($S=1$) states forming on the doped Ni sites as a result of Hund coupling. Reconciling the presence of $S=1$ moments with the mainstream theories of unconventional superconductivity is challenging \cite{Jiang2019}.         

\begin{figure*}[t]
\begin{center}
\includegraphics[angle=-90, width=0.66\columnwidth]{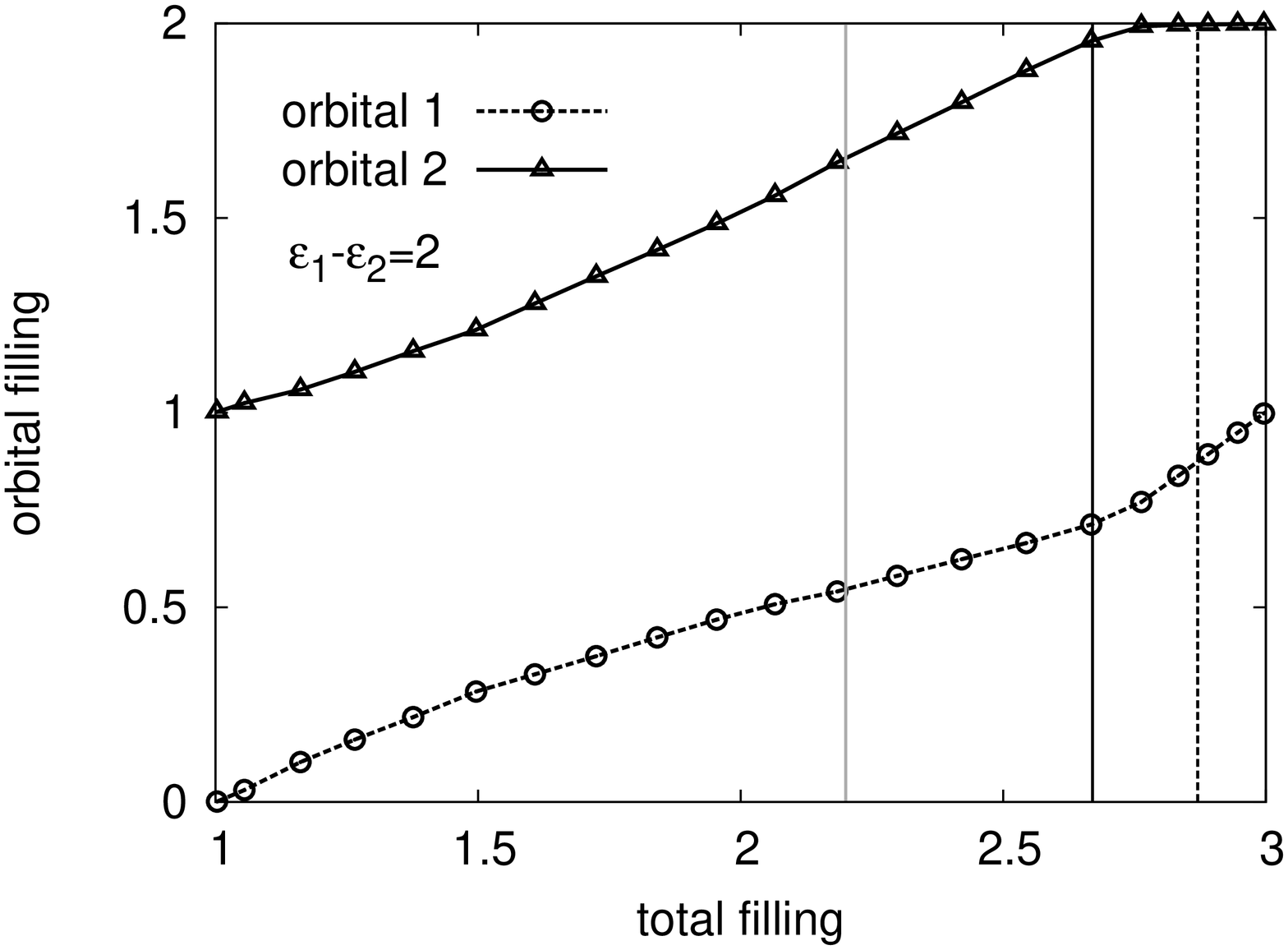}
\includegraphics[angle=-90, width=0.66\columnwidth]{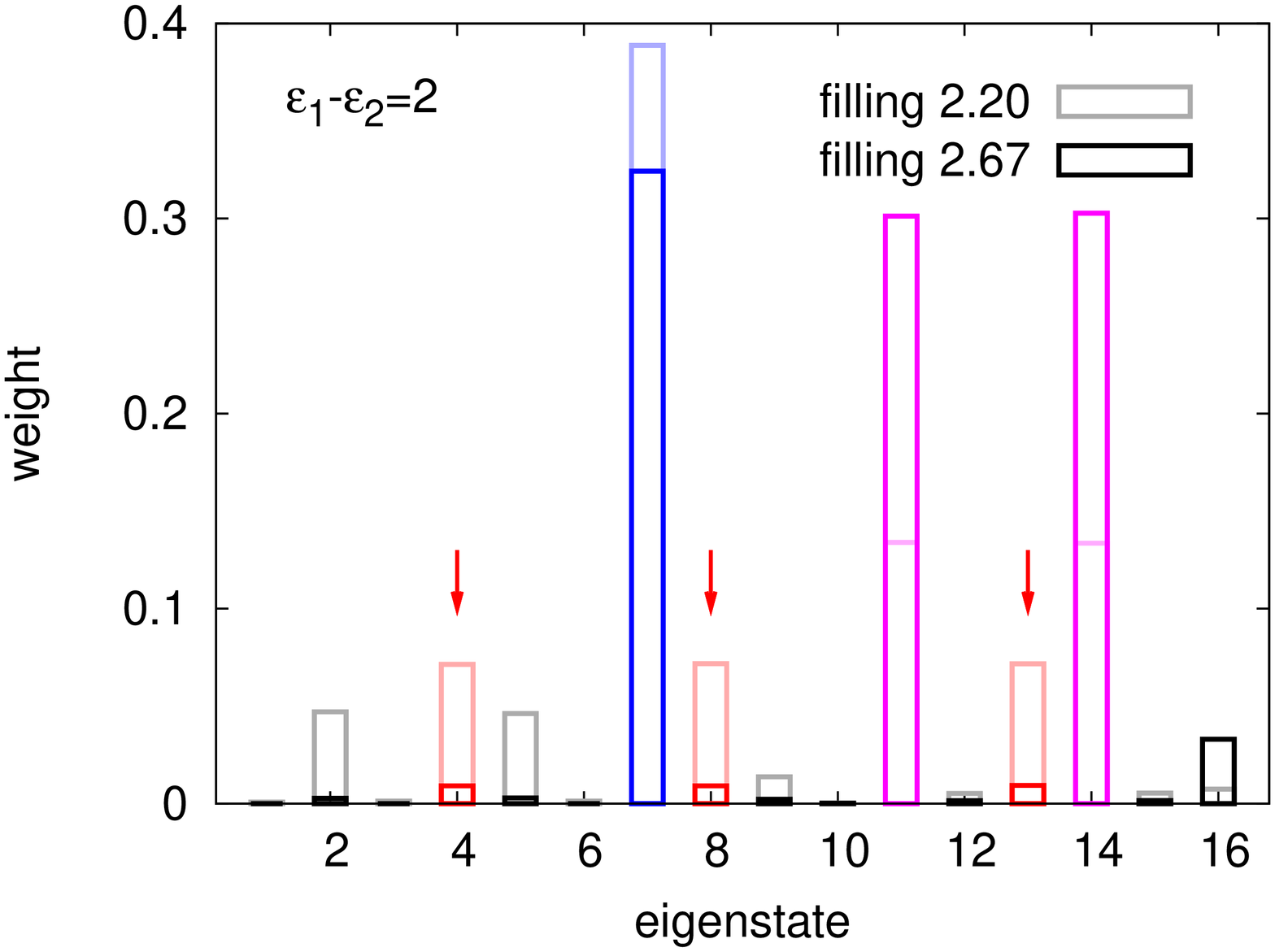}
\includegraphics[angle=-90, width=0.66\columnwidth]{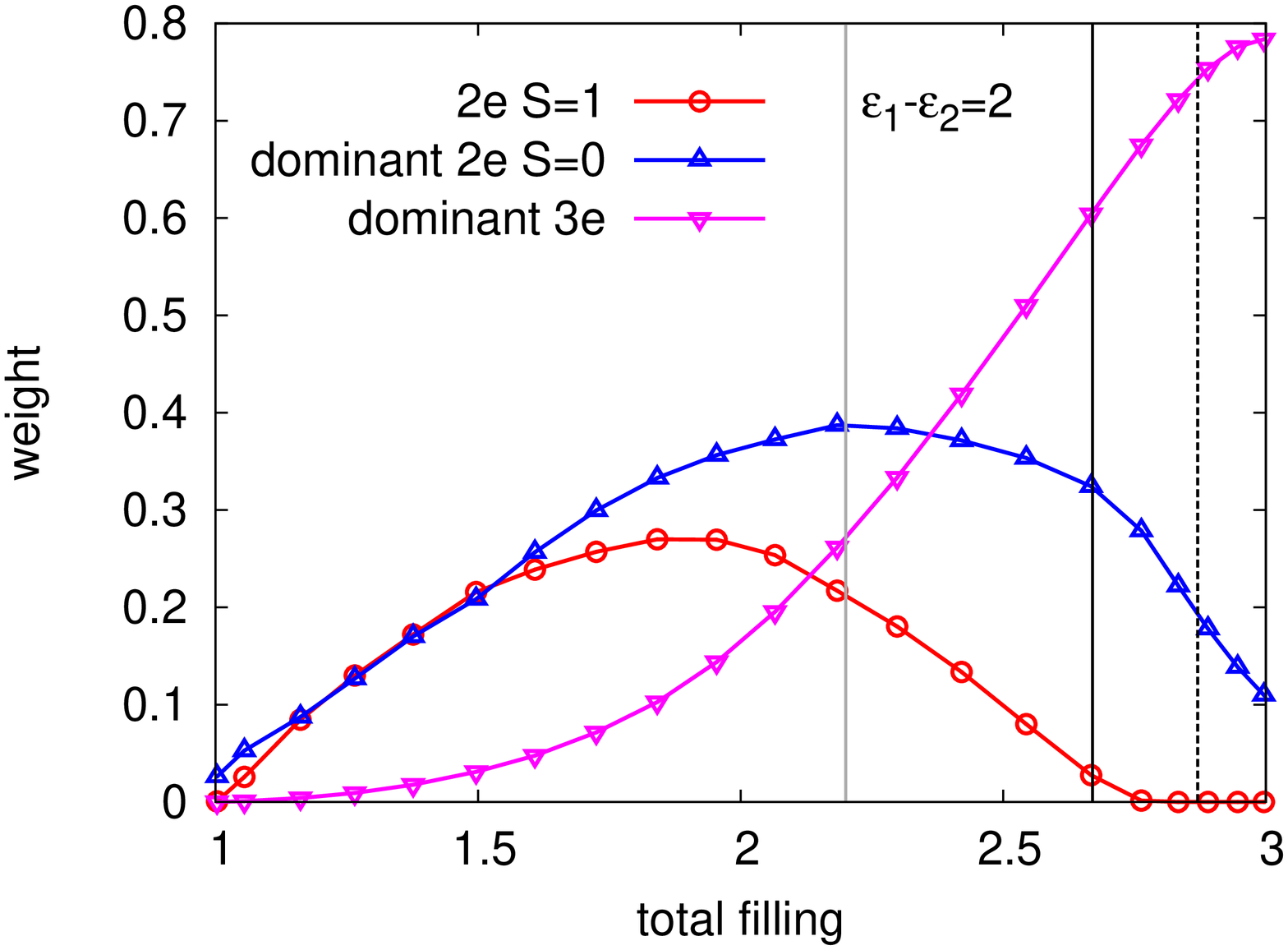}\\
\includegraphics[angle=-90, width=0.66\columnwidth]{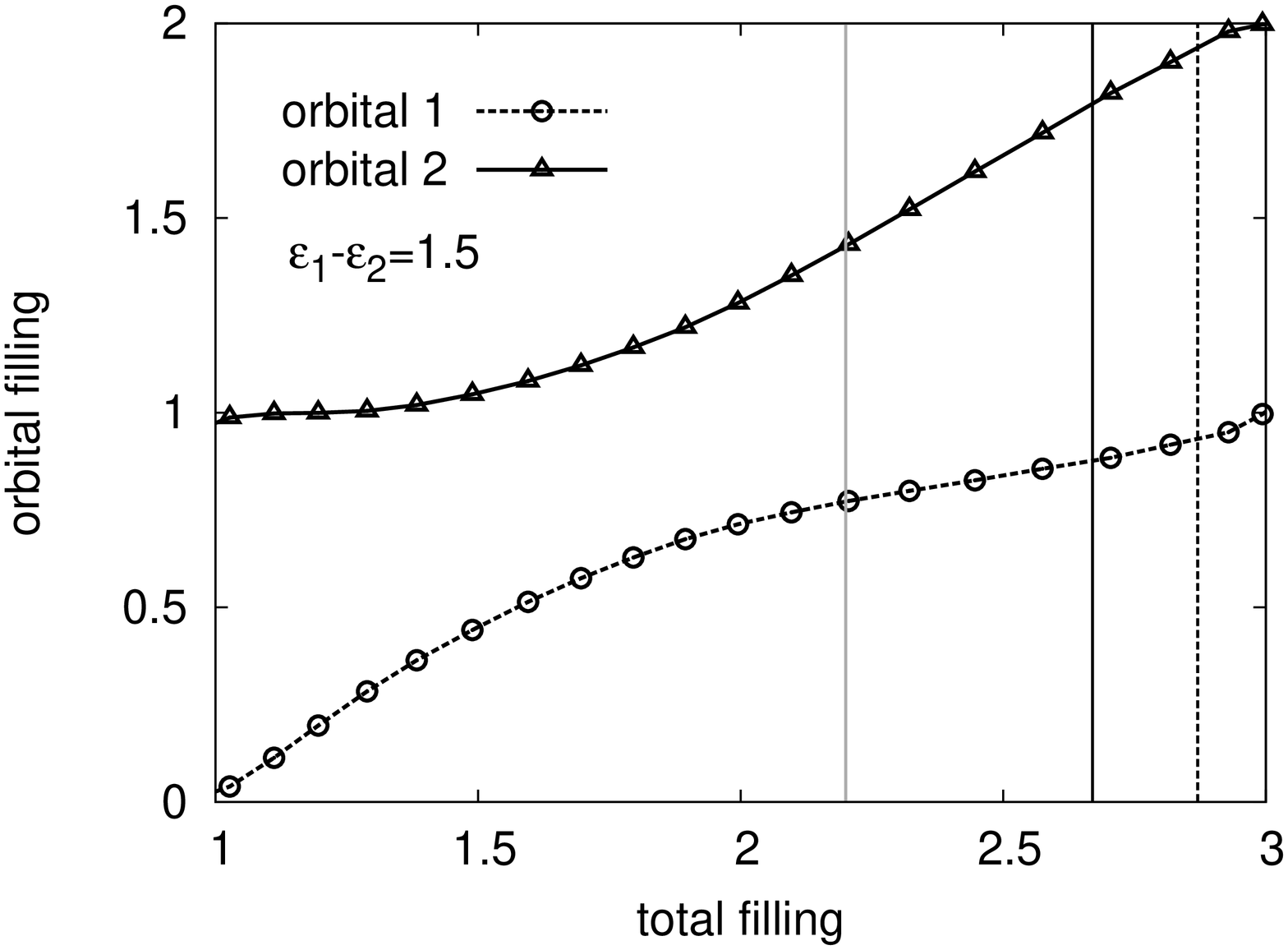}
\includegraphics[angle=-90, width=0.66\columnwidth]{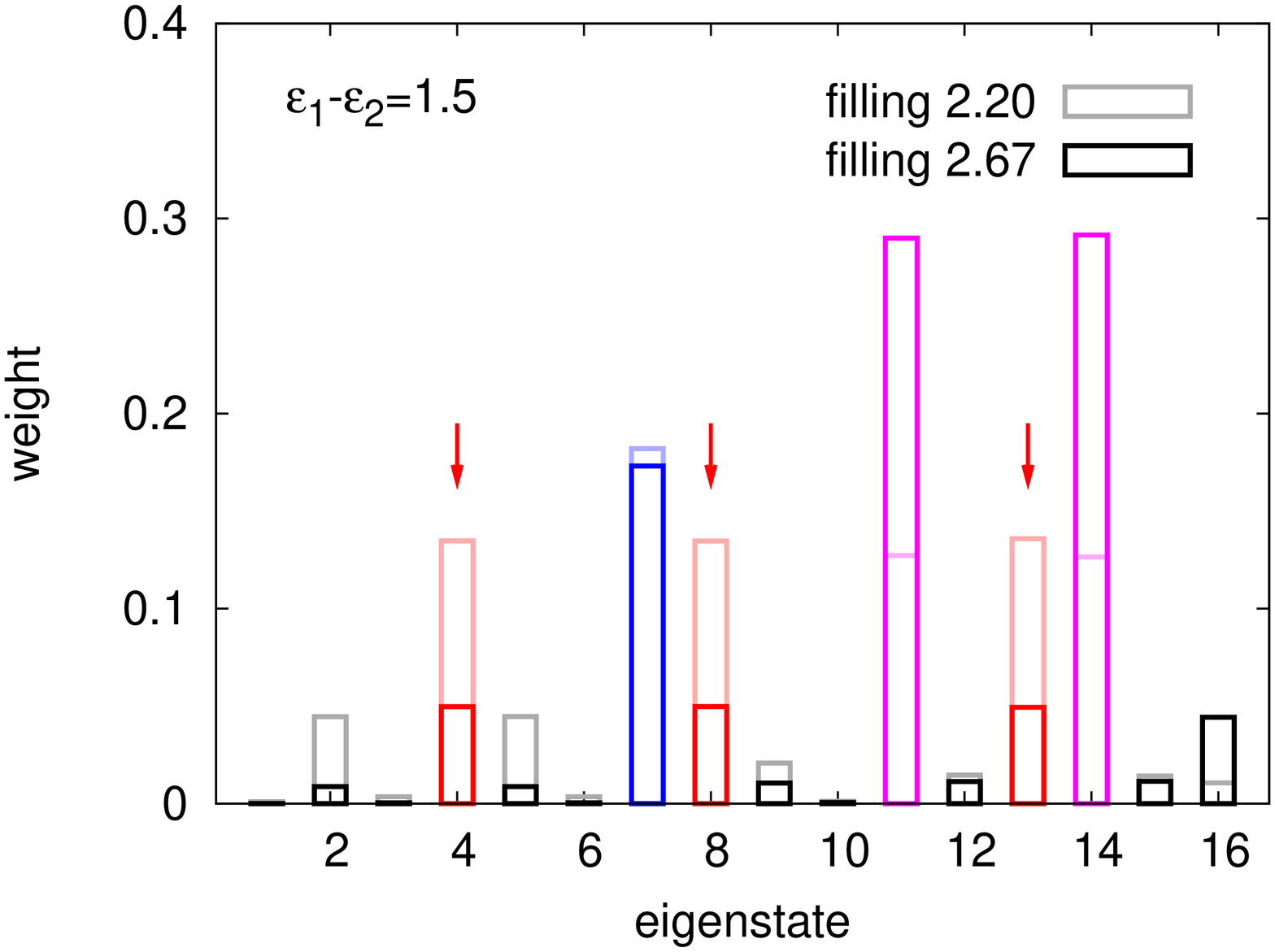}
\includegraphics[angle=-90, width=0.66\columnwidth]{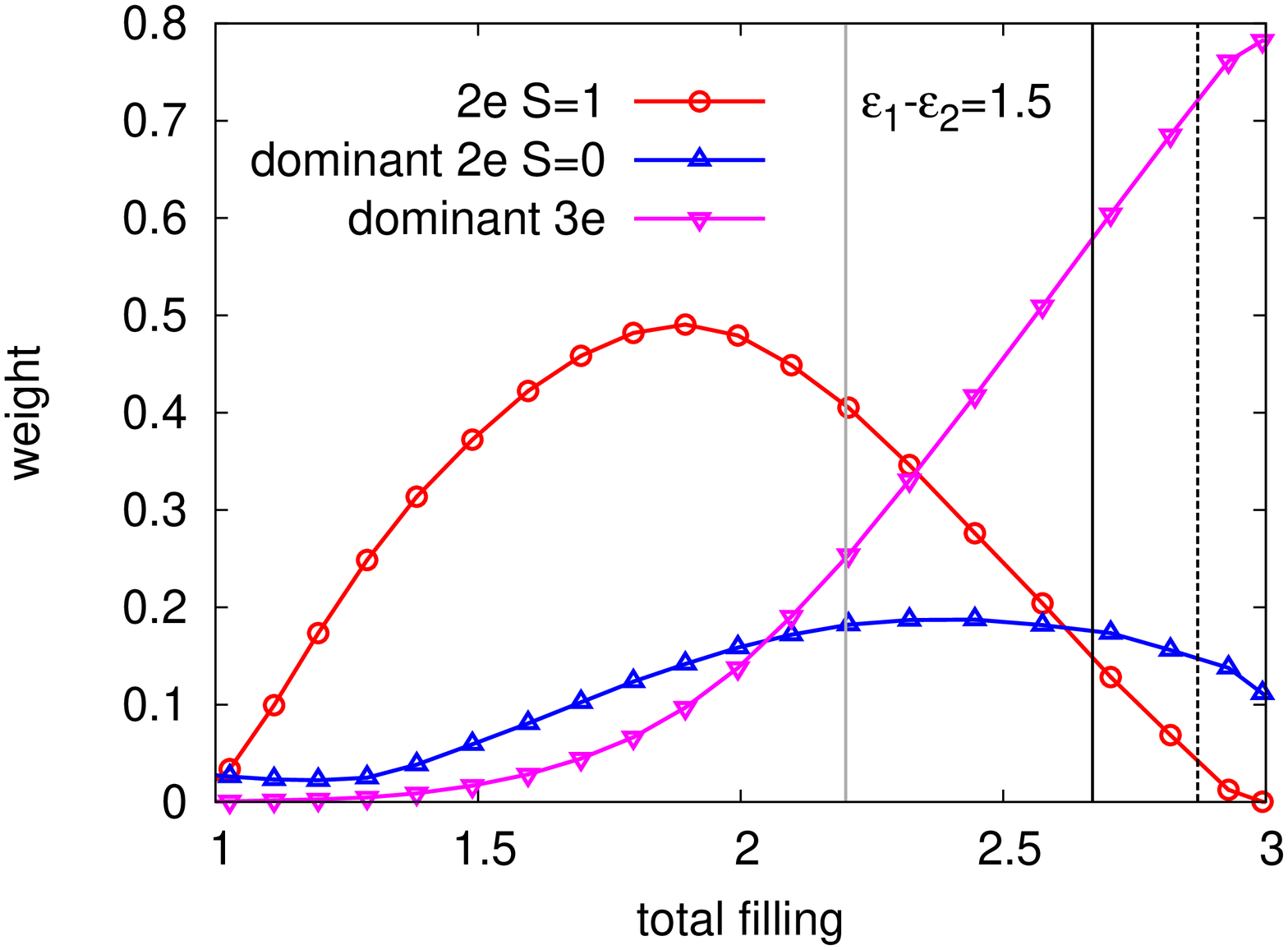}
\caption{
Left panels: Orbital filling versus total filling. The solid black vertical line indicates the experimental filling of 2.67 electrons, and the dashed black vertical line the filling of self-doped NdNiO$_2$. Middle panels: Histogram of probabilities of a given site to be in one of the 16 eigenstates of $H_\text{loc}$. The arrows mark the three triplet states. Right panels: Probability of the triplet states (red), the dominant low-spin two-electron state (blue), and the sum of the two dominant three electron states (pink) as a function of total filling.  
}
\label{fig_filling}
\end{center}
\end{figure*}

In this paper, we address these issues by considering the new nickelate high-$T_c$ superconductor from the viewpoint of the spin-freezing theory of unconventional superconductivity \cite{Hoshino2015,Werner2016}. Spin freezing \cite{Werner2008,Georges2013} refers to the formation of slowly fluctuating local moments in a physical or auxiliary multiorbital system, as a result of a physical or auxiliary Hund coupling. The spin-frozen regime extends over a finite doping range in doped Mott insulators and (in the crossover regime from spin-frozen to conventional Fermi liquid metal) results in the characteristic non-Fermi liquid behavior typically associated with the normal phase of unconventional superconductors \cite{Werner2008,Werner2018}. In Ref.~\onlinecite{Hoshino2015} we showed that there is a deep connection between spin freezing and unconventional superconductivity. Specifically, in multi-orbital Hubbard models with nonzero Hund coupling, an unconventional orbital-singlet, spin-triplet superconducting phase appears in the spin-freezing crossover regime at low temperature. The ``glue" for this superconducting state, which is most directly relevant for uranium based spin-triplet superconductors \cite{Aoki2019}, is provided by the local moment fluctuations in the spin-freezing crossover regime. It was subsequently shown that an analogous mechanism, but with enhanced local orbital fluctuations instead of spin fluctuations, explains the appearance of an unconventional spin-singlet superconducting state in multi-orbital models with negative Hund coupling \cite{Capone2009,Nomura2012}, which are relevant for the description of fulleride superconductors \cite{Steiner2016,Hoshino2017}.  The unconventional $d$-wave superconducting state in the most basic model for cuprates, the two-dimensional square-lattice single-band Hubbard model, can also be naturally understood in terms of spin freezing \cite{Werner2016}. The idea here is to map the plaquette of four sites considered in cluster dynamical mean field theory \cite{Hettler1998,Lichtenstein2000} to a pair of effective two-orbital models with large Hund coupling, through a bonding-antibonding transformation along the diagonals of the plaquette. Spin freezing in this context implies the appearance of composite high-spin moments on the diagonals of the plaquette, and the fluctuations of these moments can be argued to provide the glue for the $d$-wave superconductivity \cite{Werner2016}. This body of recent works constitutes a unified theory of unconventional superconductivity, and it is interesting to ask how the new nickelate superconductor fits into this framework.  

{\it Weak antiferromagnetic exchange.} The spin-freezing mechanism is based on {\it local} moment fluctuations. 
Antiferromagnetic correlations among the composite spins or individual spin-$\tfrac12$ moments may become important close to integer filling, but they are not essential for the pairing. Hence, the relatively high $T_c$ in the nickelate superconductor, in spite of the weak antiferromagnetic exchange coupling \cite{Jiang2019}, is not a puzzling result. 

{\it High-spin moments on the Ni site.} The calculation of the spin state and orbital occupation in multiorbital systems with Hund coupling and crystal field splittings is a nontrivial problem \cite{Werner2007,Werner2009,Suzuki2009,Kunes2014}, which however can be addressed within dynamical mean field theory (DMFT) \cite{Georges1996,Werner2006}. To get an idea about the situation in hole-doped NdNiO$_2$ we consider a two-band Hubbard model (representing the Ni $d_{x^2-y^2}$ and $d_{3z^2-r^2}$ orbitals) with approximately the bandwidths, band splittings and interaction parameters reported for the two-band model in Ref.~\onlinecite{Sakakibara2019}. For simplicity we use a Bethe lattice with a semi-circular density of states and an orbital-diagonal hybridization in the DMFT calculation. Specifically, we use the following Slater-Kanamori form of the local Hamiltonian:
\begin{align}
H_\text{loc}=&\sum_{\alpha=1,2} U n_{\alpha\uparrow} n_{\alpha\downarrow} + \sum_\sigma [U' n_{1\sigma} n_{2\bar\sigma}+(U'-J) n_{1\sigma} n_{2\sigma}]\nonumber\\
&-J(d^\dagger_{1\downarrow}d^\dagger_{2\uparrow}d_{2\downarrow}d_{1\uparrow}+d^\dagger_{2\uparrow}d^\dagger_{2\downarrow}d_{1\uparrow}d_{1\downarrow}+ \text{h.c.})\nonumber\\
&+\epsilon_1 n_1 + \epsilon_2 n_2 -\mu(n_1+n_2),
\label{eq_Hint}
\end{align}
with $U=2.6$ the intra-orbital interaction, $U'=1.3$ the inter-orbital opposite-spin interaction, $J=0.5$ the Hund coupling, $\epsilon_1=0$ ($\epsilon_2=-2$) the center of the $d_{x^2-y^2}$ ($d_{3z^2-r^2}$) band of width 3 (2), and $\mu$ the chemical potential. The unit of energy is eV, and temperature is set to $\frac{1}{50}$.

\begin{figure*}[t]
\begin{center}
\includegraphics[angle=0, width=1.3\columnwidth]{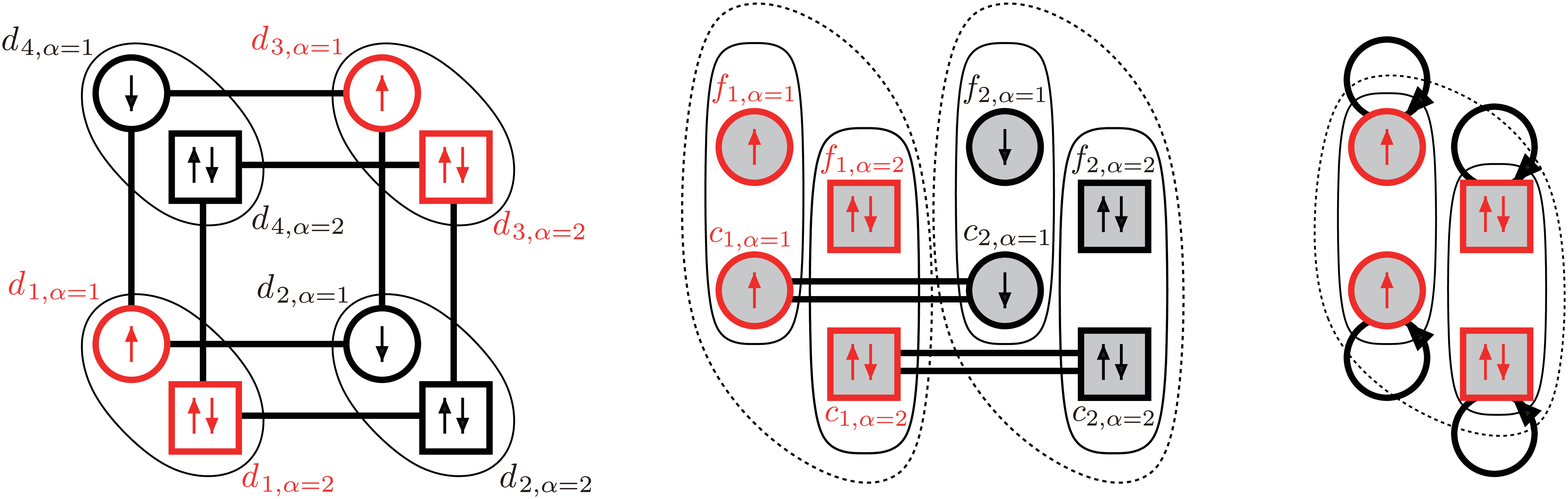}
\caption{
Left panel: Plaquette with two orbitals per site, and orbital-diagonal hopping. The upper orbital ($\alpha=1$) is represented by circles, and the lower one ($\alpha=2$) by squares. Thin ovals represent a Slater-Kanamori type interaction between the orbitals. Middle panel: auxiliary two-site four-orbital system obtained by the bonding-antibonding transformation. Here, the upper orbitals represent the bonding combination ($f$) and the lower ones the antibonding combination ($c$). Thin ovals represent a Slater-Kanamori interaction with parameters $\tilde U=\tilde U'=\tilde J=U/2$, and the dashed ovals an interorbital interaction with parameters $\hat U'=\frac{U'}{2}$, $\hat J=\frac{J}{2}$. Right panel: Hybridization structure in the auxiliary single-site four-orbital DMFT description.   
}
\label{fig_plaquette}
\end{center}
\end{figure*}

In the left panel of Fig.~\ref{fig_filling} we show the filling of orbital 1 and 2 as a function of total filling. The dashed black vertical line indicates the filling of 2.87 electrons corresponding to self-doped (by Nd 5$d$ pockes) NdNiO$_2$, while the black vertical line shows the 2.67 electron filling corresponding to Sr$_{0.2}$Nd$_{0.8}$NiO$_2$. It is evident that for $\epsilon_2-\epsilon_1=2$, the hole doping leads to a reduction of the filling in orbital 1 while orbital 2 remains essentially full down to the experimentally relevant filling. Upon further hole doping there is however a substantial drop in the occupation of the lower orbital 2, which suggests the formation of high-spin moments. Deeper insights into the relevant atomic states can be obtained from the histogram of eigenstates of $H_\text{loc}$, see middle panel. This histogram shows the probability of a given lattice site to be in one of these eigenstates. The dark colored bars correspond to the experimental filling of 2.67. The dominant state 7 is a two-electron low-spin state ($\sim$ two electrons in orbital 2), while the two subdominant states 11 and 14 are the dominant three-electron states ($\sim$ two electrons in orbital 2 and one electron in orbital 1). The two-electron triplet states correspond to the states 4, 8, 13 (highlighted with arrows) and are seen to contribute only 1\% of the weight each. This result is consistent with the picture of doped holes forming low-spin states on Ni, with an essentially empty orbital 1, and hence with a single-band Hubbard model description of Sr$_{0.2}$Nd$_{0.8}$NiO$_2$. 

In the strongly hole-doped regime, the situation is different, as the weight of the triplet states becomes significant. The light colored bars show the histogram for filling 2.2 (gray vertical line in the left panel), where the system has a 22\% probability to be in one of the triplet states.  The right panel plots the evolution of the total triplet weight (red) and the weight of the dominant two-electron $S=0$ state (blue) as a function of total filling. For comparison we also show the total weight of the two dominant three-electron states (pink). This figure demonstrates the rapid increase in the triplet weight as the filling is reduced from $2.7$ to $2$.

The doping range in which the $S=1$ states can be neglected depends on the parameters of the model, and in particular on the ratio between Hund coupling and crystal field splitting. A substantial increase in $S=1$ states can be expected once the energy cost of promoting an electron into the upper orbital becomes comparable to the gain in Hund energy. To demonstrate this, we plot in the lower panels of Fig.~\ref{fig_filling} the analogous results for a smaller crystal field splitting of $\epsilon_1-\epsilon_2=1.5$. This choice is still reasonable, especially if we consider the lower band to be a $d_{xy}$, $d_{xz}$ or $d_{yz}$ band, since these are closer to the Fermi level than the $d_{3z^2-r^2}$ band in NdNiO$_2$ \cite{Wu2019}. Now, the doping evolution of the orbital occupation is qualitatively different, in the sense that for dopings beyond a few percent, and down to a filling of about 2, most doped holes end up in the lower orbital. This is a clear indication that high-spin states are formed on the doped sites. Only for fillings below 2, when the Hund coupling becomes ineffective, does the upper orbital empty. The importance of the triplet states is directly confirmed in the histogram of eigenstates and in the right panel, which shows that the triplet weight increases almost linearly with hole doping and reaches a peak value of about 50\% near 2 electron total filling. These results show that the effect of the crystal field splitting on the spin state in our NdNiO$_2$-inspired model is quite subtle. A proper assessment of the role of the high-spin states will require a careful estimation of the interaction parameters, preferably using a fully self-consistent ab-initio scheme such as GW+DMFT \cite{Biermann2003,Nilsson2017}, and most likely also a model which includes all five $d$ orbitals. 

{\it Spin freezing.}
Since a significant fraction of doped sites in high-spin configurations is a plausible scenario, we will now investigate how the presence of $S=1$ moments affects the spin-freezing crossover behavior. The connection to spin freezing can be made by mapping a plaquette of four sites, with two orbitals per site, to a pair of auxiliary four-orbital atoms, using a bonding-antibonding transformation along the diagonals of the plaquette \cite{Werner2016}. Specifically, if the fermionic annihilation operators in the original model are denoted by $d_{i\alpha\sigma}$, and the sites of the plaquette are numbered in an anti-clockwise fashion, $i=1,2,3,4$, we define the antibonding ($c$) and bonding ($f$) operators as follows:
\begin{align}
&c_{1\alpha\sigma}=\tfrac{1}{\sqrt{2}}(d_{1\alpha\sigma}+d_{3\alpha\sigma}), \quad c_{2\alpha\sigma}=\tfrac{1}{\sqrt{2}}(d_{2\alpha\sigma}+d_{4\alpha\sigma}),\nonumber\\
&f_{1\alpha\sigma}=\tfrac{1}{\sqrt{2}}(d_{1\alpha\sigma}-d_{3\alpha\sigma}), \quad f_{2\alpha\sigma}=\tfrac{1}{\sqrt{2}}(d_{2\alpha\sigma}-d_{4\alpha\sigma}).\nonumber
\end{align}
As a result of this transformation, we obtain a 4-orbital model defined in terms of $c_{1\alpha\sigma}$ and $f_{1\alpha\sigma}$, and an analogous one defined in terms of $c_{2\alpha\sigma}$ and $f_{2\alpha\sigma}$. In Fig.~\ref{fig_plaquette} we illustrate the original two-orbital plaquette in the left panel, and the auxiliary system consisting of two four-orbital models (shaded red and black) in the middle panel. In the figure, solid lines represent a hopping $t$ between neighboring sites, and double lines a hopping $2t$. Thin solid ovals indicate a Slater-Kanamori type interaction between the encircled orbitals. In the original model (left panel), we have the interaction defined in Eq.~(\ref{eq_Hint}) between the orbitals $\alpha=1$ (circles) and $\alpha=2$ (squares). After the transformation (middle panel), we have a {\it different} Slater-Kanamori interaction between the $c$ and $f$ orbitals with the same site and same $\alpha$ index. This interaction has the unusual parameters $\tilde U=\tilde U'=\tilde J=\frac{U}{2}$, where $U$ is the original intra-orbital interaction \cite{Werner2016}. In particular, there is a very strong Hund coupling acting between the electrons in the encircled orbitals. The orbitals with different $\alpha$ (shaded circles and squares) interact with half of the original interaction parameters in a Slater-Kanamori fashion ($\hat U'=\frac{U'}{2}$, $\hat J=\frac{J}{2}$). In addition, there are correlated hopping terms involving all four flavors. For simplicity, we neglect the non-density-density interactions between the $\alpha=1$ and $\alpha=2$ orbitals in the following. The energy splitting between the $\alpha=1$ and $2$ orbitals remains unchanged under the transformation. 

Upon embedding of the plaquette into a square lattice, the cluster DMFT construction leads to a coupling of each orbital to 
hybridization functions. While the auxiliary two-site four-orbital cluster DMFT problem is completely equivalent to a four-site two-orbital cluster DMFT, it is natural to reduce the problem in the new basis to an auxiliary single-site four-orbital DMFT problem.  The corresponding interaction and hybridization structure is sketched in the right hand panel. 
(We use the modified bandwidths $W_f=2.4$, $W_c=4.2$ for $\alpha=1$ and $W_f=1.6$, $W_c=2.8$ for $\alpha=2$ \cite{Werner2016}.) This single-site treatment decouples the spin-freezing physics from the antiferromagnetic correlation effects, which involve inter-site correlations, and allows to reveal the local spin fluctuations which are relevant for superconductivity.  

\begin{figure}[t]
\begin{center}
\includegraphics[angle=0, width=0.9\columnwidth]{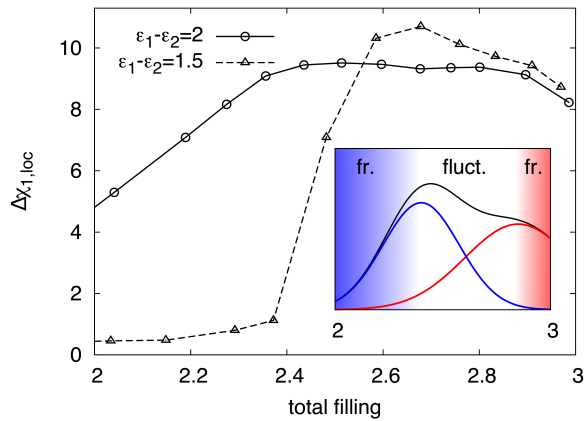}
\caption{Dynamical contribution to the local spin susceptibility in orbital 1 as a function of total filling. A large value of $\Delta\chi_{1,\text{loc}}$ near the experimentally relevant filling of 2.67 is consistent with $d$-wave pairing at low temperature. The suppression near total filling of $3$ and $2$ is due to spin freezing, which leads to a pseudo-gapped  metal state that lacks the local spin fluctuations responsible for pairing. Inset: schematic representation of the two spin-freezing crossovers associated with filling 2 and 3, showing the frozen (fr.) and fluctuating (fluct.) regions.     
}
\label{fig_deltachiloc}
\end{center}
\end{figure}

A useful quantity to analyze is the dynamical contribution to the local spin susceptibility \cite{Hoshino2015} defined as $\Delta\chi_\text{loc}=\int_0^\beta d\tau \langle S_z(\tau)S_z(0)\rangle-\beta\langle S_z(\beta/2)S_z(0)\rangle$, where the first term is the total spin susceptibility and the subtracted term represents the contribution from the frozen local moments. The spin-freezing crossover regime, with slowly fluctuating local moments, is characterized by an enhanced $\Delta\chi_\text{loc}$. Here, we focus on the spins in the upper ($\alpha=1$) orbital and compute $\Delta\chi_{\alpha=1,\text{loc}}$ using $S_z^{\alpha=1}=S_{f,z}^{\alpha=1}+S_{c,z}^{\alpha=1}$.  The results  are plotted in Fig.~\ref{fig_deltachiloc}. We find an enhancement in a broad doping range, from half-filling down to a filling of about $2.4$. In the model with smaller crystal field splitting, the fluctuations are suppressed more strongly as we approach 2 electron filling, while the magnitude of $\Delta\chi_{1,\text{loc}}$ near the experimental filling of 2.67 electrons is similar in both cases. For the interpretation of the broad peak in $\Delta\chi_{1,\text{loc}}$ it is important to realize that this hump is the result of {\it two} spin-freezing crossovers, as shown by the sketch in the inset. There is one spin-freezing crossover associated with the high-spin states formed near 3 electron filling (red), and there is a second one associated with the high-spin states formed near 2 electron filling (blue). For the interaction parameters used in this study \cite{Sakakibara2019}, the 3 electron solution is not Mott insulating \cite{footnote}, which is the reason why we do not observe a strong decrease in $\Delta\chi_{1,\text{loc}}$ near this filling. Apart from this, the behavior near 3 electron filling is completely analogous to the spin-freezing crossover in the single-band model discussed in Ref.~\onlinecite{Werner2016}. The spin-freezing behavior near 2 electron filling is more prominent and could in principle be observed already in the original single-site two-orbital DMFT solution. This crossover is associated with the formation of high-spin moments due to Hund coupling in the strongly hole-doped system, and hence the crossover from an effective single-band to a two-band picture. In the model with smaller crystal field splitting, the high-spin moments are more prominent (see Fig.~\ref{fig_filling}), and the freezing of these moments occurs at a smaller hole doping. Hence, $\Delta\chi_{1,\text{loc}}$ is strongly suppressed already at filling 2.4. In the large crystal field splitting case, where the high-spin moments never dominate the physics, the local spin fluctuations persist down to lower fillings. 

Since the enhanced $\Delta\chi_{1,\text{loc}}$ can be argued to provide the glue for $d$-wave pairing \cite{Werner2016} the consequences of Fig.~\ref{fig_deltachiloc} for superconductivity in the nickelate compounds can be summarized as follows: near 2 electron filling and (for slightly larger $U$) also near 3 electron filling the local moments freeze giving rise to a pseudo-gapped bad metal state. This state is not favorable for superconductivity and will be prone to competing magnetism or excitonic order \cite{Kunes2014,Hoshino2016}. In the experimentally relevant doping region and down to a filling which depends on the ratio between Hund coupling and crystal field splitting the system is in a spin-freezing crossover regime with enhanced local moment fluctuations. This is the non-Fermi-liquid state out of which $d$-wave superconductivity naturally emerges at low enough temperature. 

In summary, our model calculations suggest that a multiorbital description of nickelate superconductors is needed, especially in the strongly hole-doped regime, due to a subtle competition between Hund coupling and crystal field splitting. Because of the emerging multiorbital nature, the material is characterized by enhanced local spin fluctuations over a broad doping range. Nickelate superconductors are thus another family of unconventional superconductors whose physics can be naturally interpreted within the spin-freezing theory of superconductivity, in which local moment fluctuations, rather than antiferromagnetic fluctuations, induce the pairing. 
 
{\it Acknowledements} The calculations have been performed on the Beo04 cluster at the University of Fribourg, using a code based on ALPS \cite{ALPS}.

\end{document}